# Science Requirements and Trade-offs for the MOSAIC Instrument for the European ELT

C. J. Evans[1], M. Puech[2], M. Rodrigues[2], B. Barbuy[3], J.-G. Cuby[4], G. Dalton[5,6], E. Fitzsimons[1], F. Hammer[2], P. Jagourel[2], L. Kaper[7], S. L. Morris[8], T. J. Morris[8], and the MOSAIC Science Team

[1] UK Astronomy Technology Centre, Royal Observatory, Blackford Hill, Edinburgh, EH9 3HJ, UK
[2] GEPI, Observatoire de Paris, PSL Research University, CNRS, Univ. Paris Diderot, Sorbonne Paris Cité, 5 Place Jules Janssen, 92195 Meudon, France
[3] Universidade de São Paulo, IAG, Rua do Matão 1226, Cidade Universitária, São Paulo, 05508-900
[4] Aix Marseille Université, CNRS, LAM (Laboratoire d'Astrophysique de Marseille) UMR 7326, F-13388, Marseille, France
[5] Astrophysics, Department of Physics, Denys Wilkinson Building, Keble Road, Oxford, OX1 3RH, UK
[6] RALSpace, STFC Rutherford Appleton Laboratory, HSIC, Didcot, OX11 0QX, UK
[7] Astronomical Institute Anton Pannekoek, Amsterdam University, Science Park 904, 1098 XH, Amsterdam, The Netherlands
[8] Department of Physics, Durham University, South Road, Durham, DH1 3LE, UK

**ABSTRACT**

Building on the comprehensive White Paper on the scientific case for multi-object spectroscopy on the European ELT, we present the top-level instrument requirements that are being used in the Phase A design study of the MOSAIC concept. The assembled cases span the full range of E-ELT science and generally require either 'high multiplex' or 'high definition' observations to best exploit the excellent sensitivity and spatial performance of the telescope. We highlight some of the science studies that are now being used in trade-off studies to inform the capabilities of MOSAIC and its technical design.

## 1. INTRODUCTION

The scientific case for multi-object spectroscopy (MOS) on the European Extremely Large Telescope (E-ELT) has been under active development in parallel to the design of the observatory and its instrumentation. The E-ELT will be the ideal facility for optical/near-IR spectroscopic follow-up of faint sources from a diverse range of facilities, including: the *James Webb Space Telescope (JWST),* the *Euclid* and *WFIRST* missions, the Large Synoptic Survey Telescope (LSST), the Atacama Large Millimetre Array (ALMA), the *Athena* X-ray telescope, the Square Kilometre Array (SKA), and more. Indeed, we already have compelling targets from, e.g., deep imaging with the *Hubble Space Telescope (HST),* which are simply beyond the sensitivity of spectroscopy with current 8-10m class observatories.

The science case and the associated requirements that these place on the design and capabilities of an ELT-MOS have been presented in past volumes of this conference series (see Evans et al. 2012, 2014), with the comprehensive case detailed in a White Paper released early last year (Evans et al. 2015). These inputs have been used by the MOSAIC Consortium to consolidate the requirements from each case into a set of top-level requirements (TLRs) for the Phase A conceptual design of the instrument, which commenced a ~20 month study, under contract with ESO, in March 2016.

In brief, Evans et al. (2015) considered eight key science cases (SCs) to arrive at the TLRs for the Phase A study:

- SC1: 'First light' – Spectroscopy of the most distant galaxies;
- SC2: Evolution of large-scale structures;
- SC3: Mass assembly of galaxies through cosmic time;
- SC4: Active Galactic Nuclei (AGN)/Galaxy co-evolution & AGN feedback;
- SC5: Resolved stellar populations beyond the Local Group;
- SC6: Galaxy archaeology;
- SC7: Galactic Centre science;
- SC8: Planet formation in different environments.

The broad theme of the first seven is charting galaxy evolution over cosmic time, from resolved stellar populations in nearby galaxies to observations of the most distant galaxies. In addition, SC8 draws on the latest results in the quickly growing field of exoplanet research. Within these cases we identified three observational modes (see Evans et al. 2012):

- *High-definition mode (HDM):* Observations of tens of channels at fine spatial resolution, with multi-object adaptive optics (MOAO) providing high-performance AO for selected sub-fields (e.g. Rousset et al. 2010);
- *High multiplex mode (HMM):* Integrated-light (or coarsely resolved, via GLAO) observations of >100 objects.
- *Intergalactic-medium mode (IGM mode):* Parts of SC2 and 3 require spatially-extended spectroscopy at visible wavelengths (where GLAO correction is sufficient in terms of spatial resolution).

In this paper we summarise the TLRs and introduce some of the science trade studies underway to prioritise capabilities within the hardware budget of €18M, and to investigate the scientific benefit of other capabilities if a larger hardware budget were to become available. This contribution complements a number of others in these proceedings, namely an overview of the MOSAIC project (Hammer et al.), some of the design issues and trade studies underway (Rodrigues et al.), and some of the tools being used for science simulations (Puech et al.).

## 2. TOP-LEVEL REQUIREMENTS FOR MOSAIC PHASE A STUDY

In the course of work on the White Paper and in subsequent discussions ahead of the start of the Phase A study, the TLRs for each mode, as outlined in Tables 1, 2, and 3, have been agreed within the MOSAIC Consortium.

**Table 1:** Top-level requirements on the high-multiplex mode (HMM).

| Parameter | Value | Tolerance/Comments |
|---|---|---|
| Aperture | Near-IR: 0.6"; visible: 0.9" | Assumes GLAO performance in the near-IR. |
| Multiplex | 200 | Min: 200; Max: none. Takes account of cross-beam switching for sky subtraction. |
| Spec. resolving power ($R$) | 5000: visible & near-IR 15000: visible | Min requirements, at central wavelength in the relevant band. $R \geq 15{,}000$ is desirable in the near-IR. |
| Total wavelength coverage | 0.4 to 1.8 μm | Simultaneous coverage of whole range is not required. |
| Simultaneous wavelength coverage | one broad band at $R \geq 5000$ > 0.04 μm at $R \geq 15000$ | > 0.1 μm at $R = 5000$. Simultaneous coverage in each band does not need to be equal. Bluer settings will have smaller coverage. |
| Science field of view | 40 arcmin$^2$ | Min: 40 arcmin$^2$; Max: 78 arcmin$^2$ (max field of E-ELT). |

**Table 2:** Top-level requirements on the high-definition mode (HDM).

| Parameter | Value | Tolerance/Comments |
|---|---|---|
| Aperture | 2.0" × 2.0" | Values are min requirements. No trade-off in size or shape possible. |
| Multiplex | 10 | Min: 10; Max: none. |
| Spatial pixel size | 75 mas | Min: 40 mas; Max: 80 mas. |
| Ensquared energy (EE) | 30% | Min: 25%. Defined as EE in 2×2 spaxels in the *H*-band, anywhere in the MOSAIC science field of view. |
| Spec. resolving power ($R$) | 5000 | Min value. |
| Total wavelength coverage | Essential: 1.0 to 1.8 μm Desirable: 0.8 to 2.5 μm | Simultaneous coverage of whole range is not required. Trade-offs required of upper and lower limits of HDM. |
| Simultaneous wavelength coverage | One broad band at $R \geq 5000$ | Simultaneous coverage in each band does not need to be equal. Bluer settings will have smaller coverage. |
| Science field of view | 40 arcmin$^2$ | Min: 40 arcmin$^2$; Max: 78 arcmin$^2$ (max field of E-ELT). |

**Table 3:** Top-level requirements on the intergalactic-medium mode (IGM).

| Parameter | Value | Tolerance/Comments |
|---|---|---|
| Aperture | 3.0" × 3.0" | 2.0" × 2.0" as a minimum. Needs trade-off of aperture vs. multiplex. |
| Multiplex | 30 | Min: 10. |
| Spatial pixel size | 0.25" | Min: 0.2"; Max: 0.3". (May use same aperture as HMM-visible). |
| Spec. resolving power ($R$) | 5000 | Min value. 3000; Max: 5000. |
| Total wavelength coverage | Essential: 0.4 to 0.8 μm Desirable: 0.37 to 1.0 μm | Simultaneous coverage of whole range is not required. Trade study needed of expected blue performance of E-ELT cf. this requirement. |
| Simultaneous wavelength coverage | 0.38 to 0.6 μm; one broad band at λ >0.6 μm. | > 0.1 μm at $R = 3000$. |
| Science field of view | 40 arcmin$^2$ | Min: 40 arcmin$^2$; Max: 78 arcmin$^2$ (max field of E-ELT). |

Within the Phase A study we will consider these as <u>four observational modes</u>, with HMM split into HMM-NIR and HMM-Vis given the different spectrographs/treatment required for the HMM observations, providing a better systems-level fit with the near-IR HDM and visible IGM mode.

The scientific analysis toward the technical design has involved two key aspects since the start of the Phase A study:
- *Blue-visible coverage (≤ 0.4 μm):* Analysis of potential performance to quantify the competitiveness of E-ELT observations in this λ-region.
- *Coverage of K-band (in HDM):* Analysis of potential performance in the *K*-band cf. *JWST*.

The background and latest results from this work are now expanded on in the following sections.

## 3. ANALYSIS OF BLUE-VISIBLE COVERAGE

Three of the SCs advanced in the White Paper require coverage down to at least 0.40 μm, with extensions desirable further bluewards to ~0.37 μm, to give access to:
- SC2: enhanced coverage of the Lyman-α forest for IGM tomography using galaxies at $z$ ~2-2.5;
- SC3: Lyman-α and other rest-frame UV lines in high-$z$ galaxies to constrain escape fractions and feedback;
- SC6: access to important stellar lines (e.g. the CaII *H* and *K* doublet) for Galaxy archaeology.

In thinking about the specifications for the MOSAIC design, we first need to consider the performance of the telescope itself in the blue-visible region. To optimise the performance across the full visible/IR domain, the E-ELT mirrors will be coated with a combination of silver and aluminium (Ag+Al), which gives a significant throughput gain compared to bare aluminium, but has a reflectivity of only ~30% at 0.4 μm, with a sharp drop-off to shorter wavelengths.

### 3.1. End-to-end throughput estimates

For an initial estimate of the potential blue-visible throughput we needed to make several assumptions about the design/performance of the spectrographs (including collimator, corrector, dispersing element, coatings, and detector), as well as the atmospheric transmission and fibre performance (assuming a fibre-based pick-off solution for the visible channels of MOSAIC). Updating design work done previously for the OPTIMOS-EVE study (e.g., Navarro et al. 2010), we consider the potential throughputs of blue- and red-optimised spectrographs. The latter is relevant as MOSAIC will potentially have only one spectrograph design for the visible (with multiple copies to provide the multiplex), which is then used to observe selected spectral regions across the visible domain.

The assumed performances for each of the elements are detailed in the Appendix, with the contributions from the atmosphere, the latest estimates available from the telescope design and the likely performance of the fibres shown in Fig. 1, together with the estimated performance for blue- and red-optimised spectrographs, in which the assumed fibre transmission is tailored for the visible (if the fibres were chosen to optimize near-IR performance their throughput drops-off significantly at <0.6 μm). The estimated end-to-end throughputs for the blue- and red-optimised spectrographs are shown in Fig. 2. In the context of the current discussion, the pertinent results are estimated overall throughputs (T) of $T_{BLUE}$ = 7.1% (blue-optimised) and $T_{RED}$ = 3.5% (red-optimised) at 0.4 μm.

### 3.2. Consequences for blue-visible performance

A first investigation of the potential of blue-visible observations with the E-ELT is to place the above throughput estimates in the context of what is possible at present with the VLT. The primary targets for the IGM tomography case are Lyman-break galaxies at $2 \leq z \leq 2.5$, with $<r'> \sim$ 24.5 mag and $<g'-r'> \sim$ 0.25 mag (e.g., see Erb et al. 2006, and taking their *G* and *R* filters as roughly equivalent to the Sloan set). The continnum brightness in the *B*-band (3 days from new moon) in the ESO Exposure Time Calculator is 22.4 mag/arcsec$^2$, i.e., irrespective of filter transformations these observations are background dominated. Similarly, the galaxy archaeology case requires blue-visible spectroscopy of stars at the main-sequence turn-off in external galaxies, down to $V \sim$ 25 mag (cf. the *V*-band sky brightness of 21.7 mag/arcsec$^2$), so these will also be background dominated. Advanced simulations are now underway within the Phase A study to further investigate the requirements the galaxy archaeology case places on the MOSAIC design.

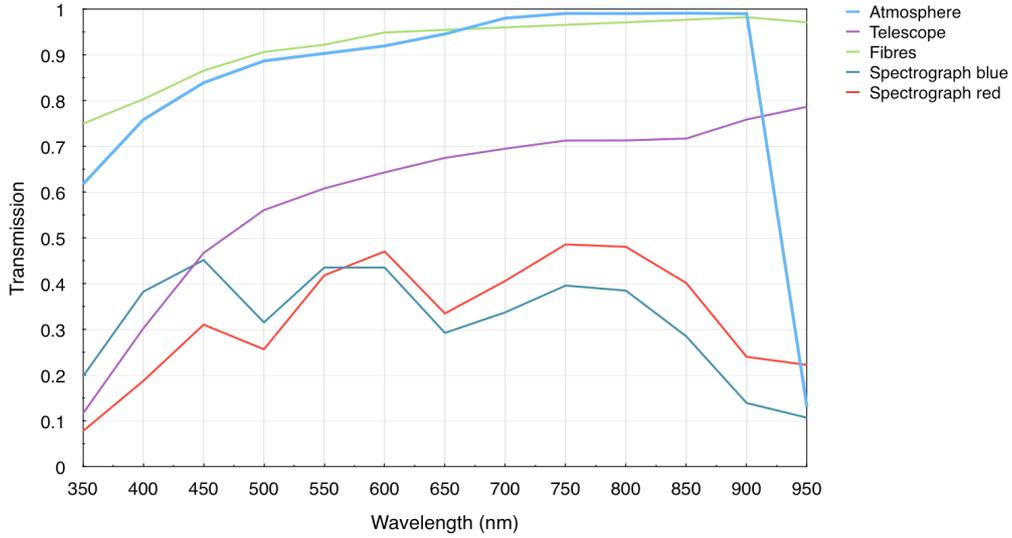

**Figure 1:** Assumed transmission of the atmosphere, telescope and fibres (see Appendix), and throughput estimates for blue- and red-optimised spectrographs.

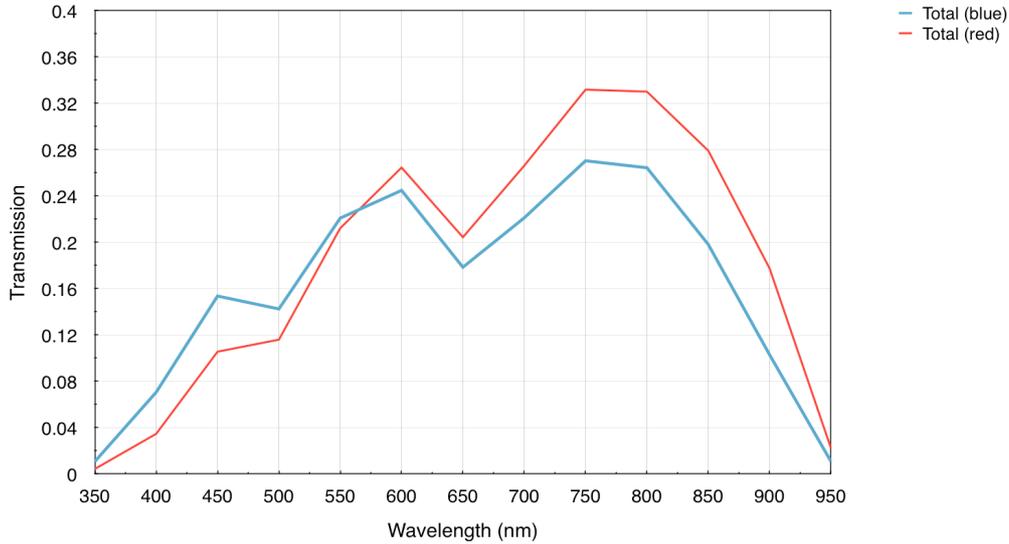

**Figure 2:** Estimated end-to-end throughputs for blue- and red-optimised spectrographs.

In the background-limited regime the signal-to-noise (S/N) of a given target (in a fixed amount of time) scales with the effective diameter (i.e. a factor of 4.8, ignoring the central obstructions), the square root of the inverse of the resolving power ($R$), and the square root of the total throughput (T):

$$S/N_{MOSAIC} / S/N_{VLT} = (38.5/8) \times \sqrt{(R_{VLT}/R_{MOSAIC})} \times \sqrt{(T_{MOSAIC}/T_{VLT})}$$

The top-level requirements for MOSAIC argue for a 0.9" on-sky aperture, which is comparable to the slit-widths considered here, so we do not consider the difference in aperture at this point. (Note that this preliminary estimate therefore assumes noise-less CCDs given that multiple fibres will be required for observations of a given target in the GLAO-corrected, high-multiplex mode.)

## 3.3. Assumptions for VLT instruments

In what follows we compare the potential MOSAIC performance with that achievable with the VLT in the blue-visible today with X-Shooter, UVES, and FLAMES-Giraffe. For the latter we consider the parameters for both the fibre-fed MEDUSA mode and the IFU mode. The relevant details assumed for these are:

- *X-Shooter:* The mean throughputs of the orders either side of 0.4 μm in the UVB arm of X-Shooter are ~9% (taken from the latest QC data). X-Shooter delivers $R$~5,000 with a 0.9" slit, which matches the high-multiplex (visible) aperture in the MOSAIC top-level requirements.

- *UVES:* The throughput at 0.4 μm is 12% (from the UVES User Manual, no atmos., wide slit). Taking into account an atmospheric throughput of ~75% (see Table 12 in the Appendix) gives an end-to-end throughput of ~9% (in good agreement the latest QC data from ESO from observations of standard stars). UVES delivers $R$ ~ 50,000 with a 1" slit.

- *FLAMES-Giraffe (Medusa fibres):* Adopting numbers from the ESO Exposure Time Calculator (ETC), observations with the LR02 set-up (spanning 0.4 μm) have a throughput of ~4%, giving a resolving power of $R$~6,000. Observations with the HR02 set-up (also spanning 0.4 μm) have a throughput of 2.25%, giving a resolving power of $R$~22,500. [We note the aperture of the Medusa fibres is 1.2", so our estimates are conservative as the fibre would see more background cf. MOSAIC.]

- *FLAMES-Giraffe (IFU):* For the IGM (IFU) case a more fitting comparison is the IFU-fed mode of FLAMES. This has similarly low throughputs as in the Medusa mode for the LR02 and HR02 settings, with a greater resolving power because of the smaller (0.52") aperture for each fibre ($R$~8,500 for LR02, $R$~35,000 for HR02). [Here we make a similar assumption as for MOSAIC that co-adding the spectra from each IFU element, in the absence of read noise, ultimately leads to the same background for the same on-sky target area.]

We note that these comparisons are somewhat oversimplified for SC2, where the target sizes are typically larger than 0.9", so there is a net gain in using (larger) IFUs than a simple slit, which would lose flux in one direction. In contrast, the FLAMES-Giraffe IFUs (~3"×2") would gather all of the flux so provide a more meaningful comparison for GLAO-corrected IFUs in MOSAIC (IGM mode).

## 3.4. MOSAIC cf. VLT

The relative S/N scalings for the MOSAIC options at 0.4 μm compared to the VLT instruments are given in Table 4. A key result is the comparison with X-Shooter, where observations with the red-optimised MOSAIC spectrograph at $R$~5,000 are only a factor of two-to-three better in terms of the delivered S/N. The increased efficiency of the blue-optimised spectrograph pushes this upwards to a factor of around four – if long exposure times were needed with the E-ELT for these observations, then this is moving into the range where VLT observations are unlikely to be feasible.

We also note that the comparisons in Table 4 are somewhat unfair in the sense that, e.g., if only $R$ = 5,000 is required, then UVES and FLAMES are effectively penalised by their greater resolving power. Neglecting read-out noise and assuming that the spectral resolution elements are sampled similarly, a UVES spectrum, for example, could be rebinned to $R$ = 5,000 to obtain better S/N. Table 5 shows the results from a similar set of comparisons, but now rebinned to the relevant resolving powers of MOSAIC.

Moving to shorter wavelengths (0.375 μm), the median throughput of X-Shooter is still fairly good (e.g. see response curves in Fig. 8 from Vernet et al. 2011). Adopting a throughput of 7.5% for X-Shooter, and interpolating between the throughput estimates for MOSAIC in Table 12 (i.e. $T_{BLUE}$ = 4.1%, $T_{RED}$ = 1.9%), the relative performances in terms of S/N are summarised in Table 6, with X-Shooter observations becoming competitive at these wavelengths for the red-optimised spectrograph.

Of course, such comparisons neglect the fact that X-Shooter and UVES are single-object instruments cf. the intended multiplex of 200 for the high-multiplex mode of MOSAIC, or the ≥10 IFUs of the IGM mode. However, *if* the good far-blue efficiency of X-Shooter/UVES could be matched in a blue-optimised, multi-object instrument on the VLT then, given the significant cost of observation time on the E-ELT, such observations might be well served by a large allocation of time on the VLT (probably not unreasonable once the E-ELT is operational) for, e.g., an optimised MOS for IGM tomography.

Table 4: Effective gain in signal-to-noise (for a given exposure time) at 0.4 μm of the blue- and red-optimised MOSAIC spectrographs compared to existing VLT instruments.

| MOSAIC (0.40 μm) | $R$ | X-Shooter ($R$~5,000) | UVES ($R$~50,000) | FLAMES (Medusa) | FLAMES (IFU) |
|---|---|---|---|---|---|
| Blue-opt ($T_{BLUE}$ = 7.1) | 5,000 | 4.3 | 13.5 | 7.0 (LR02) | 8.3 (LR02) |
| | 15,000 | 2.5 | 7.8 | 10.4 (HR02) | 12.6 (HR02) |
| Red-opt ($T_{RED}$ = 3.5) | 5,000 | 3.0 | 9.4 | 4.9 (LR02) | 5.8 (LR02) |
| | 15,000 | 1.7 | 5.4 | 7.3 (HR02) | 8.8 (HR02) |

Table 5: Effective gain in signal-to-noise (for a given exposure time) at 0.4 μm of the blue- and red-optimised MOSAIC spectrographs compared to existing VLT instruments (rebinned to the relevant resolving power of MOSAIC, ignoring read-out noise and sampling issues).

| MOSAIC (0.40 μm) | $R$ | X-Shooter (*rebinned*) | UVES (*rebinned*) | FLAMES (Med:rebinned) | FLAMES (IFU:rebinned) |
|---|---|---|---|---|---|
| Blue-opt ($T_{BLUE}$ = 7.1) | 5,000 | 4.3 | 1.4 | 5.8 (LR02) | 4.9 (LR02) |
| | 15,000 | 7.4 | 2.3 | 7.0 (HR02) | 5.4 (HR02) |
| Red-opt ($T_{RED}$ = 3.5) | 5,000 | 3.0 | 0.9 | 4.1 (LR02) | 3.4 (LR02) |
| | 15,000 | 5.2 | 1.6 | 4.9 (HR02) | 3.8 (HR02) |

Table 6: Effective gain in signal-to-noise (for a given exposure time) at 0.375 μm of the blue- and red-optimised MOSAIC spectrographs compared to X-Shooter.

| MOSAIC (0.375 μm) | $R$ | X-Shooter ($R$~5,000) |
|---|---|---|
| Blue-opt ($T_{BLUE}$ = 4.1) | 5,000 | 3.5 |
| | 15,000 | 2.0 |
| Red-opt ($T_{RED}$ = 1.9) | 5,000 | 2.4 |
| | 15,000 | 1.4 |

## 4. INCLUSION OF *K*-BAND IN THE HIGH-DEFINITION MODE

The desire for *K*-band observations with the HDM of MOSAIC is primarily driven by the extragalactic cases, and most strongly by the galaxy assembly/dynamics case (SC3, Evans et al. 2015). This case was studied in detail during the E-ELT Design Reference Mission (DRM, which ran from 2007-2009), during which more than 1000 simulations were used to explore a broad range of parameter space in terms of the physical properties of the target galaxies, as well as instrumental and observational factors (Puech et al. 2008; 2010). Here we summarise some of the relevant points which influence the potential *K*-band performance.

The DRM simulations assumed a total emissivity of 5% for the five-mirror E-ELT design with a protected Ag+Al coating (cf. an estimated emissivity for bare Al of 14%). As shown in Fig. 3, this results in a thermal background that equals the sky continuum background at ~2.15 μm, with it rising to ten times the background by 2.4 μm.

The simulated galaxies were chosen to sample the mass function at different redshifts, from dwarf (~0.1M*) to giant galaxies (10M*), investigating a range of different morphologies and kinematics, ranging from well-virialized rotating disks, to clumpy star-forming galaxies and major mergers. To save computing time, only narrow spectral ranges were simulated for each case, focussed on the relevant emission lines, i.e.: Hα in the *K*-band at *z*~2, [OII] in the *H*-band at *z*~4, and [OII] in the *K*-band at *z*~5.6 (with the latter only accessible using [OII]). Fig. 4 compares the (median emission-line) surface brightness of the simulated galaxies to the total background levels, demonstrating that the observations are background dominated in each simulation (ignoring for now the potential complication of the saturation of the night sky lines, see Rodrigues et al., these proceedings). At 2.4 μm this is dominated by the thermal background, which is mainly due to the contribution from the telescope (which the central panel of Fig. 4 shows is approx. ten times the sky signal, as expected from Fig. 3).

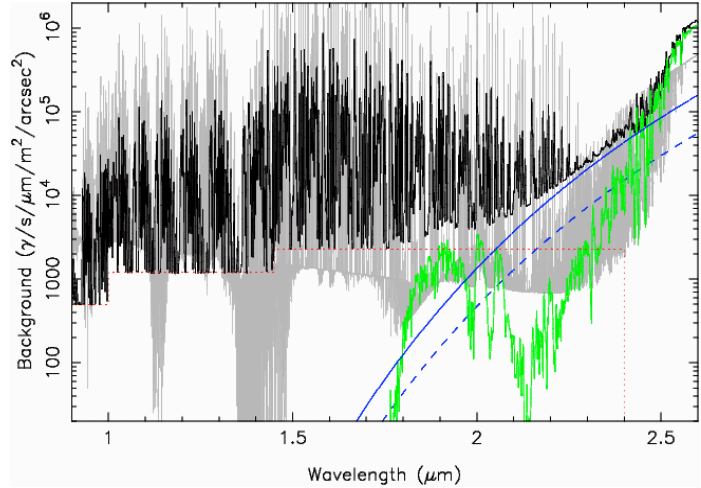

**Figure 3:** Comparison of different background sources for the E-ELT [Image credit: ESO[1]]. *Black:* Near-IR background model for Paranal-like site (airmass = 1, bare Al coating, $R$=1000); *Red (dotted):* representative continuum, assumed to be constant in each band, and which does not extend beyond the *K*-band. *Blue:* telescope thermal background assuming bare Al (solid line) and Al+Ag (dashed line) coatings. *Green:* thermal emission from the atmosphere. *Grey:* near-IR background model (at greater spectral resolution) for comparison from the Gemini ETC for Mauna Kea.

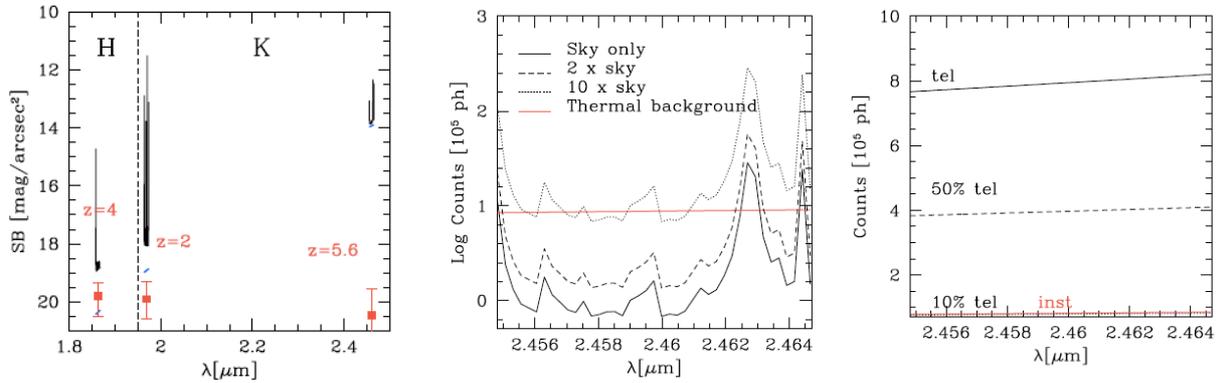

**Figure 4:** *Left:* Emission-line surface brightness of the simulated galaxies (red points, with errors bars showing the range of values for masses of 0.1 to 10M*) cf. the telescope + instrument thermal background (small blue lines) and total background (thermal + continuum + OH sky-lines; black lines). *Centre:* Contributions (in photons) to the total background in the $z = 5.6$ simulations: black lines: the sky contribution (continuum + OH lines); red line: total thermal background (telescope + instrument). *Right:* Contributions of the telescope (black) and instrument (red) to the total thermal background. [Extracted from Figs. 7 and 9 of Puech et al. (2010).]

The signal-to-noise (S/N) of observations of a simulated major merger at $z = 5.6$ (total integration time of 24 hr, $R = 5000$, 50mas/spaxel) are sufficiently affected by the background that the recovery of large-scale motions is limited, at best, to the most massive systems (see Fig. 5). This becomes even more limiting for observations of clumpy disks at $z = 5.6$ (see Fig. 6). Clumps can be resolved in relatively low-mass galaxies in simulations at $z = 4$, but this is not possible in the simulations of a 10M* galaxy at $z = 5.6$. Considering the reduced background (and atmospheric absorption bands) at 2.15 μm, we could potentially study [OII] emission in galaxies out to $z \sim 4.8$, but the S/N is still reduced by a factor of $\sqrt{2}$ by the thermal background.

The scale of the challenge for *K*-band observations of galaxies at $z = 5.6$ was highlighted by Table 6 from Puech et al. (2010), where high-quality (S/N≥10) observations of massive (~5M*) galaxies require about 16 nights per pointing, cf. less than a night to achieve similar goals in the *H*-band for galaxies *at* $z = 4$. Such simulations serve as examples of what can potentially be achieved with MOAO-corrected observations of high-*z* galaxies with the E-ELT; we now consider the impact of the different background levels on the IFU science possible with MOSAIC compared to *JWST/*NIRSpec.

---

[1] From the ESO E-ELT DRM technical data repository: https://www.eso.org/sci/facilities/eelt/science/drm/tech_data/background/

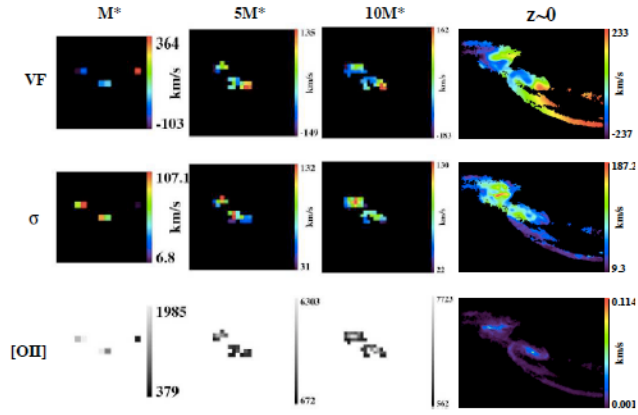

**Figure 5:** Major-merger simulations at *z* = 5.6 using MOAO. *Top to bottom:* velocity fields, velocity-dispersion maps, and emission-line maps, for different stellar masses (from 1M*, first column, to 10M*, penultimate column); lower-mass systems are not detected. The size of the images, from left to right, are: 0.71, 1.35, and 1.7 arcsec. The far-right column shows the input templates for comparison with the simulations (although the spatial scales are different). [Reproduced from Fig. 11 of Puech et al. (2010).]

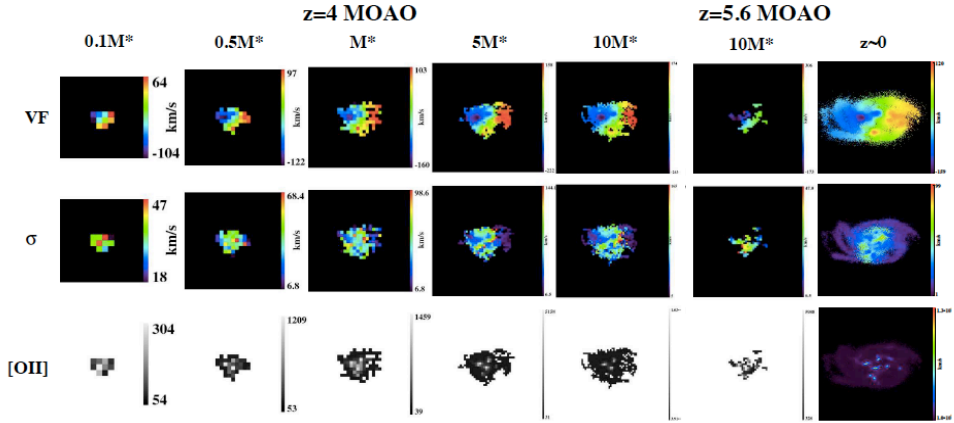

**Figure 6:** *Top to bottom:* velocity fields, velocity-dispersion maps, and emission-line maps from simulations of clumpy disks at *z* = 4 using MOAO (from 0.1M*, first column, to 10M*, fifth column; the size of the images, from left to right, are: 0.7, 1.25, 1.6, 2.8, and 3.55 arcsec), and MOAO at *z* = 5.6 (10M*; the size of the stamp is 3.0 arcsec). The far-right column shows the input templates for comparison with the simulations (although the spatial scales are different). [Adapted from Fig. 13 of Puech et al. (2010).]

## 4.1. Comparison with *JWST*/NIRSpec

The E-ELT thermal background will ultimately limit its *K*-band performance, whereas NIRSpec observations will potentially be limited by the properties of the zodiacal light and the detectors. For our comparisons we adopted the zodiacal-light model of Aldering (2001), which provides a good match to the values used in the *HST* and *JWST* exposure time calculators (Giavalisco, Sahu & Bohlin, 2002). This results in the surface brightness values and count-rate densities listed in Table 7.

**Table 7:** Zodiacal light count-rate densities from the North Ecliptic Pole model from Aldering (2001).

| Wavelength [μm] | Mag/arcsec$^2$ | Count-rate density [ph/s/μm/m$^2$/arcsec$^2$] |
|---|---|---|
| 1.25 | 22.67 | 38 |
| 1.65 | 22.79 | 26 |
| 2.15 | 23.13 | 14 |
| 2.40 | 23.35 | 10 |

We used the model values in Table 7 to estimate the relative survey speed (per spatial pixel) for point sources observed with MOSAIC cf. *JWST*/NIRSpec, as summarized in Tables 8 and 9. The parameters adopted in these calculations are:

- Sky-background values as in the ELT DRM simulations, i.e., $J$ = 1200 phot s$^{-1}$ μm$^{-1}$ m$^{-2}$ arcsec$^{-2}$, $H$ and $K$ = 2300 phot s$^{-1}$ μm$^{-1}$ m$^{-2}$ arcsec$^{-2}$, where the $J$- and $H$-band values are taken from Cuby et al. (2000).
- Following the model shown in Fig. 3, we adopt a thermal background of ten times the sky continuum at 2.4 μm, and equivalent to the sky value at 2.15 μm (with no contribution in the $J$- and $H$-bands).
- An upper limit to the dark current and readout noise for NIRSpec of 0.0041 e$^-$ s$^{-1}$ pix$^{-1}$ and 5.18 e$^-$ pix$^{-1}$, respectively (Rauscher et al. 2014). For simplicity we have adopted the same values for MOSAIC for now.
- $R$=5000 for MOSAIC cf. both NIRSpec resolving powers, i.e. $R$ = 1000 (Table 8) and $R$ = 2700 (Table 9), adopting two detector pixels per spectral resolution element in each case.
- The AO requirement for MOSAIC is 30% EE in a spatial-resolution element in the $H$-band (i.e., 150 mas, with 75 mas spatial pixels). Thus we adopt 15% pix$^{-1}$ for the $H$-band (and $J$-band), and 20% pix$^{-1}$ in the $K$-band.
- The predicted NIRSpec EE in 100 mas is ~30% (using simulated PSFs for the F140X and F110W acquisition filters made available by STScI[2]).
- Detector integration time (DIT) of 600s (background limited, but below typical timescale of substantial sky variations, and ignoring potential saturation of the sky lines for now).

From the performance estimates in Tables 8 and 9, MOSAIC is relatively competitive to NIRSpec, aside from the performance at 2.4 μm when spectral resolving power (e.g. velocity resolution) is less critical (i.e. $R$ = 1000). For the adopted parameters we note that this conclusion is valid for other DITs, (e.g. 300s, 900s), as the thermal background is the limiting factor at longer wavelengths. That said, we caution these comparisons are order-of-magnitude estimates because of the known uncertainties (e.g., differences of global throughput, the central obscurations of each telescope, spatial and temporal variations of zodiacal light, sky-subtraction strategy, delivered AO correction etc), but they serve for an initial comparison of survey efficiency.

**Table 8:** Comparison of survey efficiency from *JWST*/NIRSpec and E-ELT/MOSAIC at different wavelengths (see Figs. 3 and 4 for the different contributions to the E-ELT/MOSAIC background). Note that this assumes that the sky background can be measured and accurately subtracted by nodding the target within the IFU, otherwise the MOSAIC efficiency is reduced by a further factor of two.

|  | NIRSpec | MOSAIC 2.4 μm | MOSAIC 2.15 μm | MOSAIC 1.65 μm | MOSAIC 1.25 μm |
|---|---|---|---|---|---|
| Relative Backgrounds (incl. zod. light, sky cont., tel. thermal, detector properties) | 1 | 751 | 121 | 45 | 18 |
| Telescope diameter (m) | 6.5 | 38.5 | | | |
| Spectral resolving power ($R$) | 1000 | 5000 | | | |
| Spatial sampling (mas) | 100 | 75 | | | |
| S/N relative difference | 1 | 0.10 | 0.24 | 0.29 | 0.46 |
| Multiplex | 1 | 10 | | | |
| Survey Speed ($\propto$ S/N$^2$ × Multiplex) | 1 | 0.09 | 0.57 | 0.86 | 2.14 |
| Multiplex required to equal survey speed of NIRSpec | | 109 | 18 | 12 | 4.7 |

**Table 9:** As for Table 8, but now for $R$ = 2700 for *JWST*/NIRSpec.

|  | NIRSpec | MOSAIC 2.4 μm | MOSAIC 2.15 μm | MOSAIC 1.65 μm | MOSAIC 1.25 μm |
|---|---|---|---|---|---|
| Relative Backgrounds (incl. zod. light, sky cont., tel. thermal, detector properties) | 1 | 790 | 129 | 49 | 20 |
| Telescope diameter (m) | 6.5 | 38.5 | | | |
| Spectral resolving power ($R$) | 2700 | 5000 | | | |
| Spatial sampling (mas) | 100 | 75 | | | |
| S/N relative difference | 1 | 0.25 | 0.63 | 0.76 | 1.19 |
| Multiplex | 1 | 10 | | | |
| Survey Efficiency ($\propto$ S/N$^2$ × Multiplex) | 1 | 0.64 | 3.92 | 5.76 | 14.2 |
| Multiplex required to equal survey speed of NIRSpec | | 16 | 2.5 | 1.7 | 0.7 |

---

[2] http://www.stsci.edu/~mperrin/software/psf_library/

## 4.2. Discussion

Our initial conclusions are that the significant thermal background of the E-ELT means that MOSAIC will struggle to be competitive with observing targets one at a time with NIRSpec at 2.4 μm, but that it is otherwise relatively compelling in terms of survey speed. We also stress that the velocity/spatial resolution from MOSAIC will be better for investigation of the spatial structures and the dynamics of high-$z$ galaxies (SC2). In short, there are important science cases relating to the topic of galaxy formation and evolution which can only be satisfied with $K$-band observations. *JWST* will likely address aspects of these via a future IFU programme, but $K$-band observations with MOSAIC would provide better spatial/spectral resolution to investigate high-$z$ systems, in a moderately competitive amount of time.

With MOSAIC likely being *the* MOS on the E-ELT for several years, there is a more general argument for inclusion of the $K$-band in terms of parameter space for future discoveries. It is hard to pre-judge what the most compelling targets will be in the late 2020s, and we do not know what, e.g., ALMA or *JWST* might discover that requires spectroscopic follow-up by then. For instance, deep imaging from *JWST* will provide legacy data and spectroscopic targets beyond the timescales of NIRSpec operations, and which will require long integrations and a multiplexed instrument to provide the required sample sizes (cf. HARMONI). We note that such programmes with MOSAIC will still be demanding (e.g. the estimated exposure times from Puech et al. 2010), entailing large 'Public Survey'-like efforts.

However, given the current hardware budget available for MOSAIC (and even assuming the consortium can raise additional funds), inclusion of the $K$-band is particularly challenging. For instance, the poor performance of optical fibres in the $K$-band precludes the use of common fibre-fed spectrographs for the high-definition and high-multiplex modes. This points to a slicer-based design for the high-definition channels, which was studied within the context of the EAGLE Phase A study (Cuby et al. 2010). We are now revisiting the potential costs of such a design to investigate its financial feasibility (including updated costs for near-IR arrays and micro-deformable mirrors) as well as the technical challenges (e.g. the near-IR spectrographs would need to be mounted on the rotating mass of the instrument).

If inclusion of the $K$-band is not possible within the available budget, the primary science cases that will suffer are:
- *Galaxy kinematics at $z > 4$ (via [OII])* – progress expected from *JWST*/NIRSpec, but MOSAIC would provide better spatial/velocity resolution in a (reasonably) competitive amount of time, and for larger samples given the fixed lifetime of *JWST*.
- *Metallicity gradients out to $z \sim 4$* – such studies are needed to provide important constraints on feedback processes, and without the $K$-band they will be limited to $z \sim 2.5$, missing the important era before the peak of cosmic star-formation/AGNs. *JWST* will be competitive, but again limited by fixed lifetime.
- *Lyα escape fraction* – feasible with Hβ alone, although not optimal.
- *First-light: HeII/CIII] at $z \sim 10$-12* – hard to assess at present, and partly depends on the nature and number of candidates from *JWST* imaging. Initial spectroscopy will probably be obtained from *JWST* (albeit at lower $R$) and HARMONI spectroscopy might suffice if the source densities are low.

Inclusion of the $K$-band for the stellar cases (e.g. SCs 5 and 6) is desirable/optimal, but not essential. Equally, for studies of the region around the Galactic Centre (SC7) and in studies of young star-forming regions, the coarse-sampling mode of HARMONI could be used to map the most compelling pointings.

## 5. SUMMARY

The E-ELT will be the world's largest optical/IR telescope when complete, so we will aim for the largest possible discovery space when designing MOSAIC, balanced by the technical feasibility and cost. Following the considerable work over the past years in developing the science case for MOSAIC, we have recently consolidated the instrument requirements from this broad range of cases to a firm set of top-level requirements (Tables 1, 2, and 3).

Analysis for the trades detailed in Sections 3 and 4 is still ongoing, and will be revised/updated as the design work progresses during Phase A, and to also take into account relevant updates from the telescope (e.g. latest estimates of the thermal background). Nonetheless, our initial conclusions include:
- MOSAIC observations at 0.37-0.40 μm are compelling cf. current VLT instrumentation, although we comment that a large campaign with a blue-optimised MOS on the VLT could be competitive for some science cases. At this stage in Phase A we are continuing to investigate the performance of potential designs in the blue, keeping

- a careful eye on the estimated end-to-end throughputs to ensure that they remain competitive. In particular, if the spectrograph is (necessarily) optimised for the red-visible, and the throughputs estimated in the Appendix are not realised, then the case for coverage ≤ 0.40 μm would be significantly weakened.

- Inclusion of the *K*-band in a high-definition (IFU) mode would open-up some exciting areas scientifically, and is sufficiently competitive with spectroscopy with *JWST* such that this will still be a compelling case after the (limited) lifetime of the mission. Indeed, access to longer wavelengths will likely become even more important with time given the expectation of new discoveries with *JWST,* ALMA, *Euclid,* and other future facilities. However, building on the work done in the past conceptual studies (EAGLE, OPTIMOS-EVE), it is clear that inclusion of *K*-band will be very difficult within the current budget (even if including significant contributions from consortium partners). If included, such an instrument would almost certainly not satisfy many of the other TLRs (visible coverage, high-multiplex mode, GLAO-fed IFUs, etc).

In addition to these points, we have also been investigating the potential issue of the sky emission lines saturating in E-ELT observations. Our analysis has found that it is not possible to obtain background-limited observations (rather than observations limited by detector read noise) without the sky lines saturating in the near-IR bands (*J, H, K*); for further details see Rodrigues et al. and Hammer et al. (both in these proceedings). An option to militate against such effects of saturation could be to window the arrays to effectively mask out the sky lines (e.g. Bezawada & Ives, 2006) and this is under further study.

Future work in the Phase A study includes further development of the above issues, combined with other trade-offs on the capabilities of the instrument linked directly to the science cases, such as the definition of the 'cut-off' wavelength between the visible and near-IR spectrographs, which will partly be driven by the predicted performances of the relevant arrays (and AO correction) near the CaII triplet at ~0.85μm.

Lastly, we note that the different modes under study for MOSAIC (high multiplex, high definition) will require different 'pick-offs' to select targets in the E-ELT focal plane, analogous to the single-object and IFU modes which feed the FLAMES-Giraffe spectrograph on the VLT. Within such an architecture one could envisage a fibre-link/relay to feed the HIRES spectrograph (see Marconi et al. these proceedings) to enable parallel observations (as per the fibre-feed to UVES from FLAMES). For example, deep MOS observations of faint IGM sight-lines to Lyman-break galaxies could be complemented with simultaneous HIRES observations of a sight-line to a bright quasar. Although not currently part of our baseline plans, such an option is something that we are exploring with the HIRES team. While such a set-up (and the necessary observing strategies) may not be suitable for the majority of programmes, given the operational costs of the E-ELT, even a small number of 'Large Programmes' observed in parallel could provide a huge saving in terms of lifetime operational costs.

# APPENDIX: ASSUMPTIONS FOR BLUE-VISIBLE PERFORMANCE ESTIMATES

**Table 10:** Throughput estimates for blue-optimised spectrograph.

| Wavelength (nm) | Collimator | Corrector | VPH | Camera coating | Camera throughput | Detector | Total |
|---|---|---|---|---|---|---|---|
| 370 | 92 % | 94 % | 45.6 % | 98 % | 73 % | 70 % | 19.7 % |
| 400 | 92 % | 94 % | 74.5 % | 99 % | 75 % | 80 % | 38.3 & |
| 450 | 92 % | 94 % | 82.8 % | 99 % | 75 % | 85 % | 45.2 % |
| 500 | 92 % | 94 % | 57.0 % | 99 % | 76 % | 85 % | 31.6 % |
| 550 | 92 % | 94 % | 83.7 % | 99 % | 76 % | 80 % | 43.5 % |
| 600 | 92 % | 94 % | 83.7 % | 99 % | 76 % | 80 % | 43.5 % |
| 650 | 92 % | 94 % | 57.6 % | 99 % | 76 % | 78 % | 29.3 % |
| 700 | 92 % | 94 % | 69.2 % | 99 % | 76 % | 75 % | 33.8 % |
| 750 | 92 % | 94 % | 84.5 % | 99 % | 76 % | 72 % | 39.6 % |
| 800 | 92 % | 94 % | 84.5 % | 99 % | 76 % | 70 % | 38.5 % |
| 850 | 92 % | 94 % | 73.0 % | 99 % | 76 % | 60 % | 28.5 % |
| 900 | 92 % | 94 % | 48.0 % | 98 % | 76 % | 45 % | 13.9 % |
| 950 | 98 % | 94 % | 55.7 % | 98 % | 76 % | 28 % | 10.7 % |

**Table 11:** Throughput estimates for red-optimised spectrograph.

| Wavelength (nm) | Collimator | Corrector | VPH | Camera coating | Camera throughput | Detector | Total |
|---|---|---|---|---|---|---|---|
| 370 | 92 % | 94 % | 45.6 % | 97 % | 55 % | 37 % | 7.8 % |
| 400 | 92 % | 94 % | 74.5 % | 99 % | 66 % | 45 % | 18.8 % |
| 450 | 92 % | 94 % | 82.8 % | 99 % | 73 % | 60 % | 31.1 % |
| 500 | 92 % | 94 % | 57.0 % | 99 % | 73 % | 72 % | 25.7 % |
| 550 | 92 % | 94 % | 83.7 % | 99 % | 73 % | 80 % | 41.8 % |
| 600 | 92 % | 94 % | 83.7 % | 99 % | 73 % | 90 % | 47.1 % |
| 650 | 92 % | 94 % | 57.6 % | 99 % | 73 % | 93 % | 33.5 % |
| 700 | 92 % | 94 % | 69.2 % | 99 % | 73 % | 94 % | 40.6 % |
| 750 | 92 % | 94 % | 84.5 % | 99 % | 73 % | 92 % | 48.6 % |
| 800 | 92 % | 94 % | 84.5 % | 99 % | 73 % | 91 % | 48.1 % |
| 850 | 92 % | 94 % | 73.0 % | 99 % | 73 % | 88 % | 40.1 % |
| 900 | 92 % | 94 % | 48.0 % | 98 % | 73 % | 80 % | 24.0 % |
| 950 | 98 % | 94 % | 55.7 % | 98 % | 73 % | 60 % | 22.3 % |

**Table 12:** End-to-end throughput estimates for blue- and red-optimised spectrographs.

| Wavelength (nm) | Atmosphere | Telescope | Fibres | Blue spectrograph | Red spectrograph | Total th'put (blue) | Total th'put (red) |
|---|---|---|---|---|---|---|---|
| 370 | 61.8 % | 11.6 % | 75.0 % | 19.8 % | 7.8 % | 1.1 % | 0.4 % |
| 400 | 75.9 % | 30.2 % | 80.4 % | 38.3 % | 18.8 % | 7.1 % | 3.5 % |
| 450 | 84.0 % | 46.7 % | 86.6 % | 45.2 % | 31.1 % | 15.4 % | 10.6 % |
| 500 | 88.7 % | 56.1 % | 90.7 % | 31.6 % | 25.7 % | 14.2 % | 11.6 % |
| 550 | 90.4 % | 60.8 % | 92.3 % | 43.5 % | 41.8 % | 22.1 % | 21.2 % |
| 600 | 92.0 % | 64.4 % | 95.0 % | 43.5 % | 47.1 % | 24.5 % | 26.5 % |
| 650 | 94.6 % | 67.5 % | 95.5 % | 29.3 % | 33.5 % | 17.8 % | 20.4 % |
| 700 | 98.1 % | 69.5 % | 96.1 % | 33.8 % | 40.6 % | 22.1 % | 26.6 % |
| 750 | 99.1 % | 71.3 % | 96.6 % | 39.6 % | 48.6 % | 27.0 % | 33.2 % |
| 800 | 99.1 % | 71.3 & | 97.2 % | 38.5 % | 48.1 % | 26.4 % | 33.0 % |
| 850 | 99.2 % | 71.8 % | 97.7 % | 28.5 % | 40.1 % | 19.8 % | 27.9 % |
| 900 | 99.0 % | 75.9 % | 98.3 % | 13.9 % | 24.0 % | 10.3 % | 17.7 % |
| 950 | 13.1 % | 78.7 % | 97.2 % | 10.7 % | 22.3 % | 1.1 % | 2.2 % |

Note that if both blue- and red-optimised spectrographs were included in the design a dichroic would also potentially be required, which would impact on the expected performance.